# Transformation devices with optical nihility media and reduced realizations

Lin Xu[1*], Qiannan Wu[2], Yangyang Zhou[3] and Huanyang Chen[3†]

[1] Key Laboratory of Opto-Electronic Information Acquisition and Manipulation of Ministry of Education & Institute of Physical Science and Information Technology, Anhui University, Hefei 230601, China

[2] School of Science, North University of China, Taiyuan, Shanxi, 030051, China

[3] Institute of Electromagnetics and Acoustics and Department of Electronic Science, Xiamen University, Xiamen 361005, China

**Abstract:** Starting from optical nihility media (ONM), we design several intriguing devices with transformation optics method in two dimensions, such as a wave splitter, a concave lens, a field rotator, a concentrator and an invisibility cloak. The extreme anisotropic property of ONM hinders the fabrication of these devices, which could be effectively realized by simplified materials with Fabry-Pérot resonances (FPs) at discrete frequencies. Moreover, we propose a reduced version of simplified materials with FPs to construct a concentrator and a rotator, which is feasible in experimental fabrications. The simulations of total scattering cross sections confirm their functionalities.

**I. INTRODUCTION**

In the past decade, researchers have found that the linear response parameters of materials can be equivalently considered as the curvature of the geometry property for electromagnetic (EM) wave propagation[1-4], which originates from the form invariance of Maxwell's equations under coordinate transformation and analogizes to Einstein's theory of general relativity [5]. Based on such equivalent perspectives between material and geometry, transformation optics (TO) [6-12] was proposed as a theoretical method to design versatile optical devices and control the flow of electromagnetic waves, such as invisibility cloaks [1, 2, 13], carpet cloaks[14-16], concentrators[17-20], field rotators[21-23], EM wormholes[24] and so on. Moreover, the method of transformation optics were also introduced to manipulate other classical waves, such as acoustic waves[25], surface water waves[26, 27], elastic waves[28, 29] and thermal dynamics[30-33].

However, TO-based devices are usual required complicated response of materials which is described by electromagnetic parameters (such as permittivity tensor $\varepsilon$ and permeability tensor $\mu$). Those materials are hardly founded in native environment. Originally, the reduced invisibility cloak was achieved by split ring resonances (SSRs) in microwaves [13]. The structure of SRRs is a kind of typical metamaterials whose unit cells depend on local resonances. The EM parameters of metamaterials can dramatically change near the resonant frequencies. After that, metamaterials were widely used to fabricate TO-based devices, such as carpet cloaks[15], field rotators[23] and so on.

More recently, photonic crystals for EM waves near at Dirac-like point was found to be like zero index materials[34], which can achieve invisibility effect as well. Such effective zero index stemmed from the collective behavior of ordered photonic crystal, which is a non-local effect. In addition, photonic crystal can also serve as effective medium to construct TO-based concentrator [35].

Nihility media with $\varepsilon=\mu=0$ [36] is a kind of metamaterials with a number of applications[37]. On the perspective of transformation optics, the definition of nihility media was generalized as transformation media derived from volumeless geometrical elements [38]. Later on, the structure of subwavelength apertures in metal was proposed to effectively achieve generalized nihility media which is named "optic-null medium" [39]. For simplification and unification, we would like to call such generalized nihility media as optical nihility media (ONM). Radiation cancellation effect and the hyperlensing effect were observed in microwave at single frequency [39]. However, the ONM is extreme anisotropic in EM parameters, which is quite hard to work at broadband frequencies.

In this paper, we firstly revisit ONM from extension coordinate transformation of volumeless geometry elements in Sec. II. Then, we design several intriguing devices with ONM in Sec. III, such as a wave splitter, a concave lens, a field rotator, a concentrator and an invisibility cloak. Furthermore, we consider that the functionalities of ONM can be effectively achieved by using simplified materials with Fabry-Pérot resonances (FPs) [18, 20] for transverse-electric(TE) EM waves. Moreover, we proposed the reduced realization of ONM by using dielectric and thin metal plate in Sec. IV, which are used to construct a concentrator and a field rotator. The total scattering cross section demonstrates their functionalities in Sec. V. We finally give a conclusion in Sec. VI.

## II. The extension of volumeless geometry elements and the ONMs.

A volumeless geometry elements in one dimensional (1D), two dimensional (2D) and three dimensional (3D) space are a point (P), a line (L) and a surface (S), as shown in the first column with purple color in Table 1, respectively. The additional counterparts and dashed lines (in blue) represent small deviation $\Delta x$ of those volumeless elements in one direction. When the deviation $\Delta x$ of is zero, the purple part and the blue part are the same to denote a volumeless geometry element. Starting from a volumeless geometry elements as well as their deviation in virtual space, for a point for example, its extension coordinate transformation in 1D is a line, while its extension in 2D is a circular region, and in 3D a sphere, which are shown in second row of Table 1. Those extended geometry elements in different dimensions are all called P-type ONM, which are similar to Ref. [38]. Dashed lines are mapped to extension paths with thin blue lines in P-type ONM. The extension coordinate transformation from one point to infinity many points will result in singularities of P-type ONM in physical space. The EM parameters of P-type ONM in 3D are $\varepsilon=\mu=0$ [38], which is the zero-index materials. Moreover, the P-type ONM in 1D and 2D can be treated as subspace of P-type ONM in 3D.

A line can be extended to a rectangular region in 2D or a cylinder in 3D as shown in the third row of Table 1. Its corresponding transformation media is called L-type ONM. The EM parameters of P-type ONM in 3D are $\varepsilon=\mu=\mathrm{diag}[1,1,0]$ [38]. The L-type ONM in 2D can be considered as cross section of L-type ONM in 3D.

In the fourth row of Table 1, the expansion of a plane in 3D virtual space is a cuboid volume. Its corresponding transformation media is called S-type ONM. The EM parameters of S-type ONM in 3D are $\varepsilon=\mu=\mathrm{diag}[\infty,0,0]$ [38].

There are two important properties of ONM. One is that light waves can only propagate along the extension paths. The second property is that a phase accumulation of light wave along the extension paths is zero. In the following part, we mainly talk about L-type ONM in 2D. However, an alternative treatment is that L-type ONM is the cross section of S-type ONM.

TAB. 1 (color online) The extension of volumeless geometry elements and the corresponding ONM. The first column is volumeless geometry elements. Dashed lines are mapped to extension paths with thin blue lines in ONM.

### III. Several intriguing devices designed by ONM.

In 2D space, the extension of a line and L-type ONM can be used to design intriguing optical devices.

## A. Splitter

Consider that a line in 2D extends to rectangular L-type ONM as shown in Fig 1(a) on the left. Let's divide it with two black lines symmetrically such that the triangle with purple line and black lines is an equicrural triangle. Then we separate the originate rectangular ONM into three parts as shown in Fig 1(a) on the right. Such a configuration can serve as a wave splitter. The simulation with an incident Gaussian beam from the left demonstrates the splitting effect with L-type ONM, as shown in Fig. 1(b). All the simulation results in this paper are performed by the finite element solver COMSOL Multiphysics.

## B. Concave lens

The shape of L-type ONM can be designed at will. In Fig. 1(c), the blue boundary of L-type ONM is set to be a semi-circle. Then we obtain a concave lens designed by L-type ONM. Because there is no phase accumulation from the purple line to the blue semi-circle, the Gaussian beam focus at the center of semi-circle after passing though the concave lens as shown in Fig. 1(d).

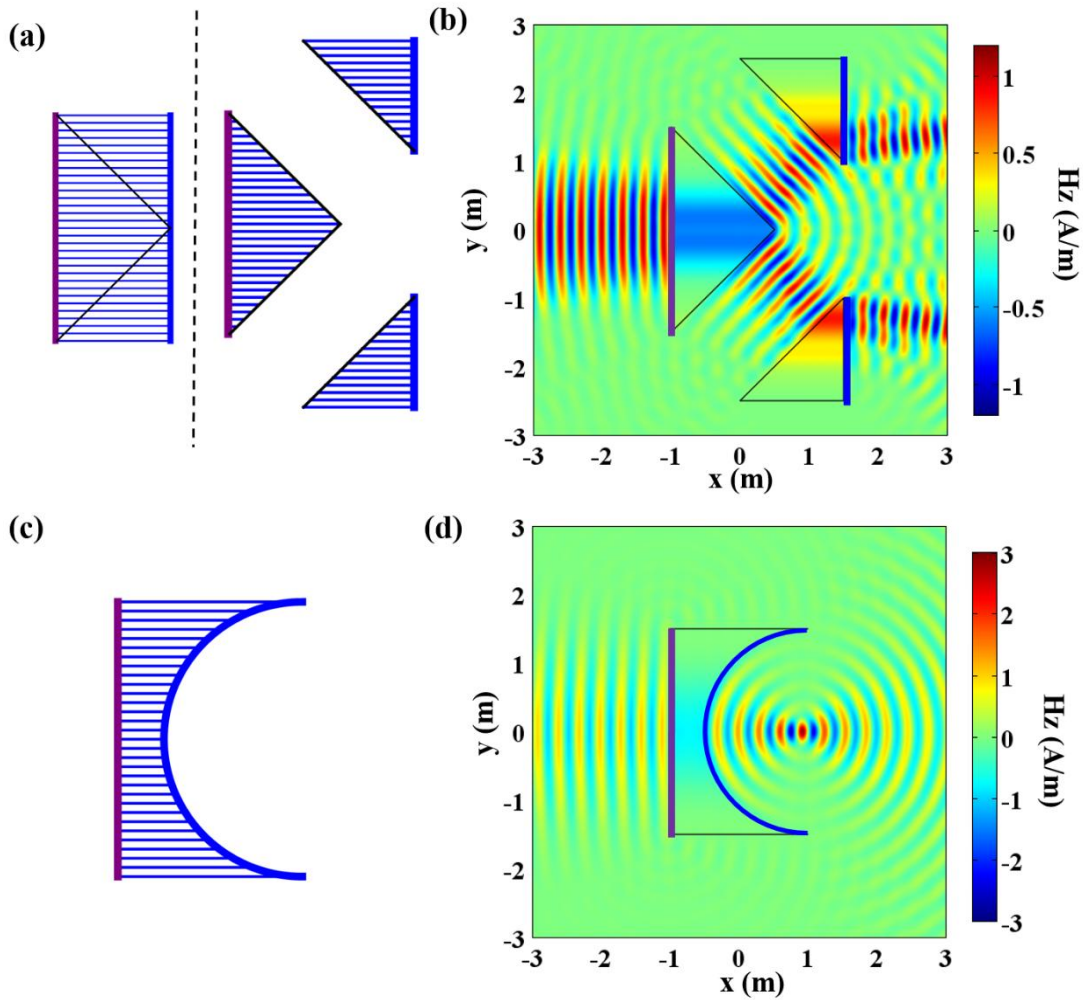

FIG. 1 (color online) (a) The design of a splitter with L-type ONM. (b) Hz field pattern of the splitter with L-type ONM. (c) The design of a concave lens with L-type ONM. (b) Hz field pattern of the concave lens with L-type ONM.

## C. Field rotator

In the previous design, the topology of ONM is the same as a point. In Fig. 2(a), we consider a volumeless element of a purple circle, it can be extended to the blue circle along curved thin lines, which forms a concentric ring region filled with ONM. The extension paths of each point in a volumeless element can be arbitrarily chosen as long as those paths do not intersect with each other. The topology of a circle is quite different from a point, it separates the whole 2D space into two disconnected regions. If we choose the extension path to be rotation linearly with the radius which is written as $\theta' = \theta + \Delta\theta \cdot (r - r_1)/(r_2 - r_1)$, we obtain L-type ONM in the concentric ring region. Furthermore, we compress the region with purple circle on the left to the region with blue circle under coordinate transformation $r' = r \cdot r_1 / r_2$. We then design a rotator with ONM, whose rotation effect under a Gaussian beam is shown in Fig 2(c).

## D. Concentrator

Based on the above rotator, if we choose the extension path to be the radius, then we can design a circular concentrator with ONM as shown in Fig. 2(d). The property of such a concentrator have been systematically investigated in Ref. [18]. Here we demonstrate that such a concentrator can be viewed as a typical case made of L-type ONM. Moreover, we can also design other concentrator with a square shape as shown in Fig. 2(e).

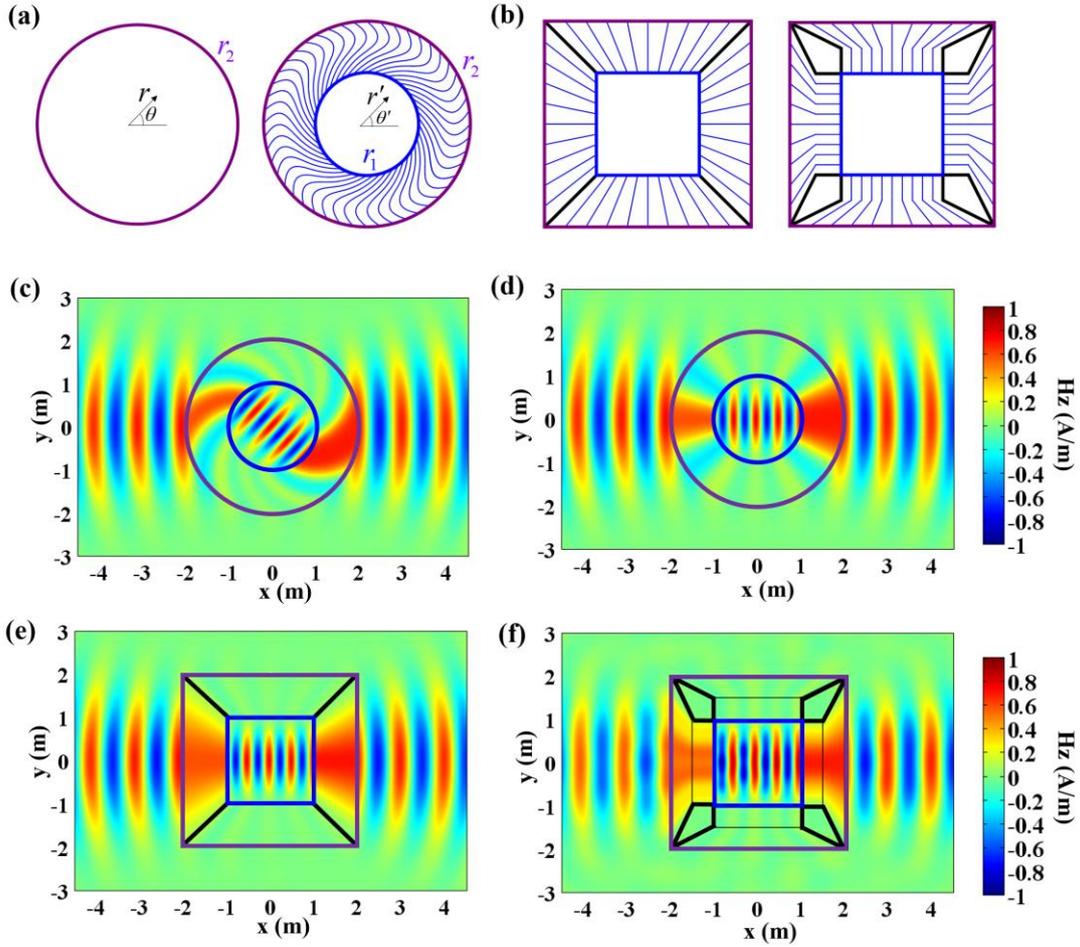

FIG. 2 (color online) (a) Extension of a circle along arbitrary paths. (b) The design of a square invisibility cloak with L-type ONM. (c) Hz field pattern of the field rotator with L-type ONM. (d) Hz field pattern of a circular concentrator with L-type ONM. (e) Hz field pattern of a square concentrator with L-type ONM. (f) Hz field pattern of a square invisibility cloak with L-type ONM. The invisible region are four quadrangles bounded with black lines.

### E. Invisibility cloak

As shown in Fig. 2(b) on the left, four thick black extension paths are highlighted in the square concentrator. In principle, those four thick black lines are invisible because this square shaped L-type ONM represent optical void. If we further extend them to four quadrangles bounded with black lines as shown in Fig. 2(b) on the right, we obtain an invisibility cloak with four invisible quadrangles. Its invisibility effect is demonstrated with an incident Gaussian beam in Fig. 2(f), where the four quadrangles are set to be metal.

**IV. Simplified materials with FPs and reduced version for L-type ONM.**

We have designed several intriguing devices with L-type ONM. How can we realize it? In the early days, the ONM was proposed and achieved by a compensated bilayer consist of positive and negative refractive index. However, the negative refractive index medium is usually obtained by local resonances [40-42], whose functionalities are compromised by the absorption. In this section, we consider that the L-type ONM can effective achieved by simplified materials with FPs. The simplified material with FPs have already been discussed in Ref. [18]. Here, we revisit its main ideal and propose a reduced version for experimental fabrications.

As we have mentioned above the two important properties of ONM. The first one is that light flows along the extension path. The second one is that there is no phase accumulation along extension path. To attain these two properties, we can reduce the second property to be the phase accumulation along extension path is integer times of $2\pi$. Such reduction of phase accumulation is also referred as FPs in Ref. [18].

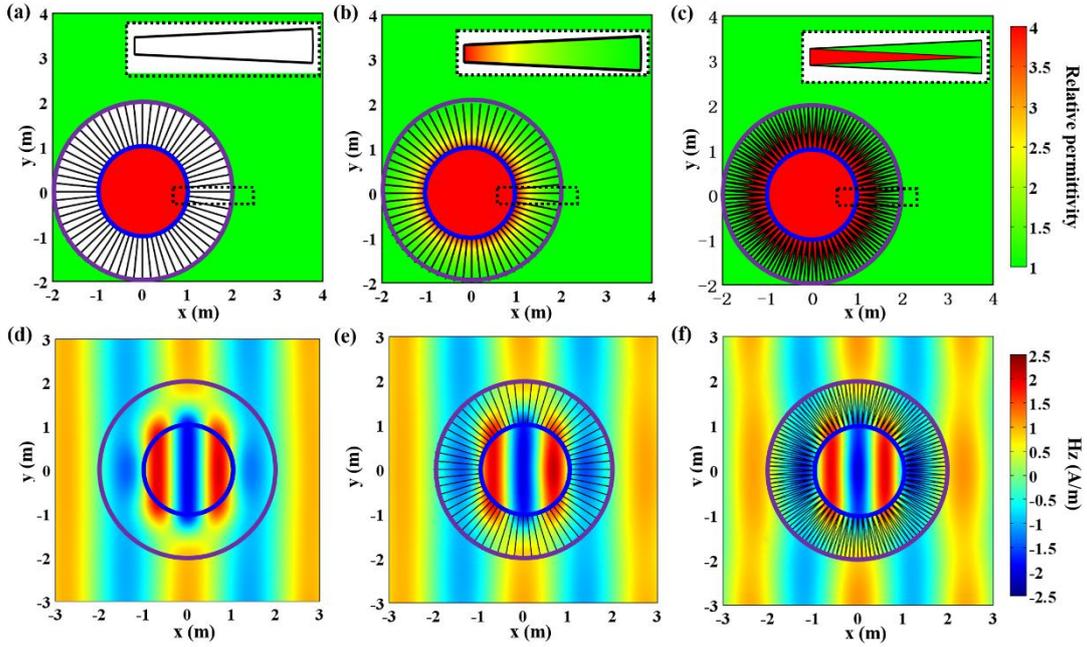

FIG. 3 (color online) (a) A concentrator of simplified material parameters with FPs. (b) A concentrator of effective gradient dielectric profiles with thin metal plates. (c) A concentrator of reduced realization with alternative dielectric and thin metal plates. (d) Hz field pattern of the concentrator of simplified material parameters with FPs. (e) Hz field pattern of the concentrator of effective gradient dielectric profiles with thin metal plates. (f) Hz field pattern of the concentrator of reduced realization with alternative dielectric and thin metal plates.

Let's consider a concrete example of concentrator. In Fig. 2(d), the ideal concentrator with ONM can collect all energy impinging at the green circle to the inner region bounded with blue circle. It is obvious that there is no phase accumulation of EM waves along the radial directions in the concentric ring region filled with L-type ONM. The L-type ONM can be replaced with simplified materials, as shown in Fig. 3(a). The main property of the simplified material parameters is the component along the radial directions of EM parameter tensors is infinity, while the components along other directions is continuous and finite. Therefore, EM waves can only propagate along the radial directions. The optical path along the radial directions in the simplified materials is $s_1 = \int_{r_1}^{r_2} \sqrt{\varepsilon} \cdot dr$. At FPs, namely the frequency $f$ satisfies

$2s_1 = l_1 \cdot c / f$ with an integer $l_1$ and the speed of light in vacuum $c$, simplified materials acts as L-type ONM. In Fig 3(b), we show that the simplified materials can effectively achieved by inserting plates made of metal (denoted with black lines) in a gradient dielectric profile, where TM polarized EM waves cannot penetrate into them, but TE waves can. As shown in Fig. 3(d), we plot the magnetic field along z direction (Hz) at FPs with $l_1 = 1$. It has been proved that the concentration effect is not perfect but still very good in such simplified materials [20]. Fig. 3(e) shows the magnetic field for the effective gradient dielectric profiles inserted with 64 pieces of metal. Each unit cell is bounded with metal and its dielectric constant is ranging from 1 to 4 with $\varepsilon = 4/r^2$. Since there is no cut-off frequency for TE-polarized EM waves in each metal waveguide, the wavelength could be simply much larger than the thickness of each unit cell for the effective medium theoretical limit. Due to the thickness of metallic plates and the simulation error, the working wavelength in Fig. 3(e) is slight shifted from FPs with $l_1 = 1$ to 1.05 compared with Fig. 3(d).

In the effective gradient dielectric profiles with metallic plates, the dielectric constant changes continuously with radius. It is not easy to fabricate such materials in experiments. In a recent study, such a method is used to concentrate the energy of water surface waves, where effective continuous refractive index profiles can be tuned by changing the depth of the water surface [26]. Here in optics, we insert each unit cell with equicrural triangle made of uniform material with its dielectric constant 4, as shown in the inset of Fig. 3(c). Therefore, we obtain a reduced version of simplified materials with a gear-shaped dielectric and metals. The effective refractive index can effectively change from 1 to 4 along the radial directions in each unit cell. It should not be exact the same profile as Fig. 3(b). By numerical calculations, we find that there is concentration and invisibility effect nearby FPs with $l_1 = 1.2$ as shown in Fig. 3(f).

To further demonstrate simplified materials with FPs and reduced version for L-type ONM, we also design a field rotator. The simplified materials are shown in Fig. 4(a). The shape of bended extension path is obtained by the rotation mapping from that of the metal plates in Fig. 3(a), where the rotation angle is set to be $\Delta\theta = 0.5\ rad.$. The optical path along the bended extension path is $s_2 = \int_{r_1}^{r_2} \sqrt{\varepsilon} \cdot dr$. At FPs, namely the frequency $f$ satisfies $2s_2 = l_2 \cdot c / f$ with an integer $l_2$ and the speed of light $c$, there should be the rotation effect. In the numerical calculations, the Hz field pattern of the simplified materials made of L-type ONM is shown in Fig. 4(d) at FPs with $l_2 = 1.05$. The configuration of effective gradient dielectric profiles with metallic plates of the rotator is shown in Fig. 4(b). There are 64 pieces inserted in gradient dielectric profiles. The unit cell is shown in the inset of Fig. 4(b). The related Hz field pattern is shown in Fig. 4(e) at FPs with $l_2 = 0.9$. The deviation of $l_2$ from 1 mainly results from the thickness of the unit cell, where the effective material parameters deviate a bit from the analytical form. By the rotation mapping, we obtain the reduced version made of bended gears and metals in Fig. 4(c) from that of Fig. 3(c). The Hz field pattern of reduced version of the rotator is shown in Fig. 4(f) at FPs with $l_2 = 0.95$, which also exhibits a little of frequency shift.

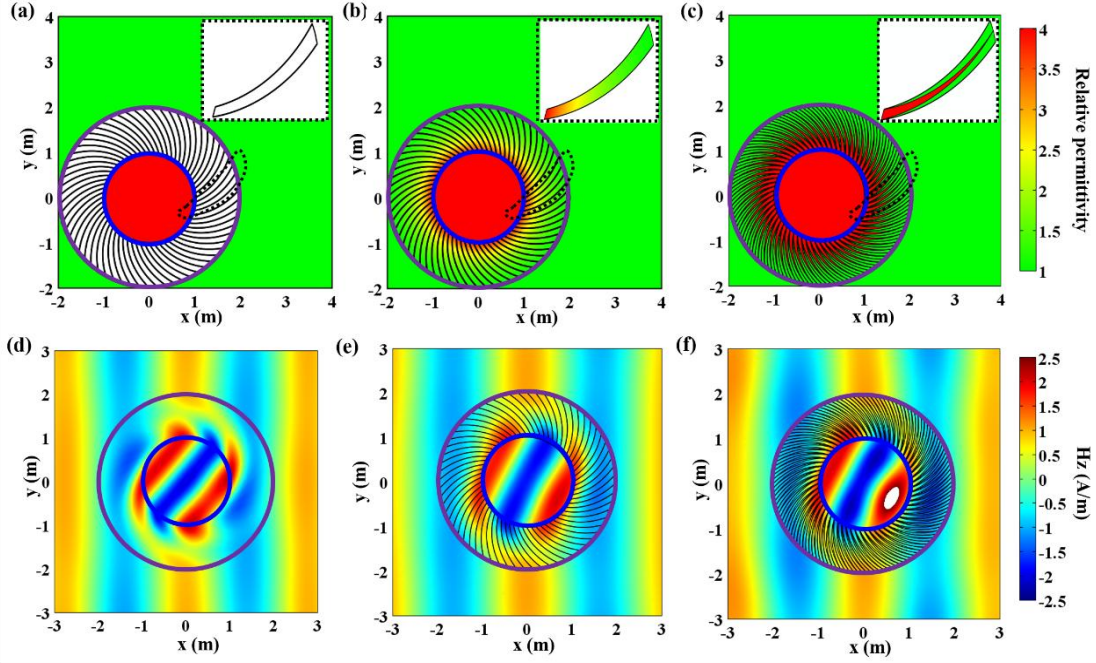

FIG. 4 (color online) (a) Field rotator of simplified materials with FPs. (b) Field rotator of effective gradient dielectric profiles with thin metal plates. (c) Field rotator of reduced version with bending dielectric and thin metal plates. (d) Hz field pattern of field rotator of simplified materials with FPs. (e) Hz field pattern of field rotator of effective gradient dielectric profiles with thin metal plates. (f) Hz field pattern of field rotator of reduced version with bending dielectric and thin metal plates.

## V. Total scattering cross sections.

To further illustrate the property of reduced versions of L-type ONM. We integrate the scattering Hz in far field as the total scattering cross sections. The integration radius is ten times the radius of bare scatter with a dielectric constant 2. The total scattering cross sections of the concentrator with simplified materials is shown in red curve in Fig. 5(a). The dips occur at FPs. Its effective concentrator is denoted with blue curve which is similar with the ideal red one. However, there is bigger scattering for the concentrator of reduced version, where the dips with green curve shift a little, comparing to the previous two curves. It is because the effective material parameters are no longer the analytical form.

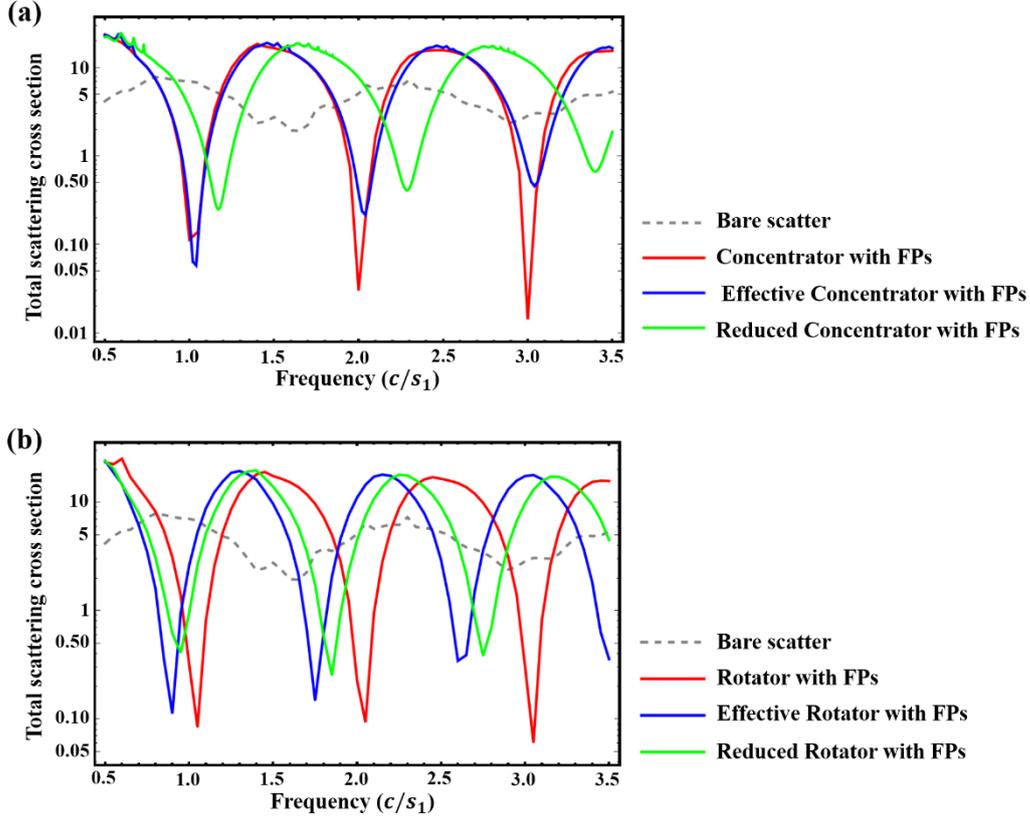

FIG. 5 (color online) Total scattering cross sections for (a) concentrators and (b) rotators.

As for the rotators, the total scattering cross sections for simplified materials, effective version and reduced version are shown in Fig. 5(b) with red, blue and green curves, respectively. The shift of the blue curve from the red curve of rotators is quite large comparing to that of the concentrator. This is because of effective medium theory is not that exact due to the bending geometry. There is also deviation for the reduced version with the green curve, where the configuration is not only bending but also with gradient thickness, hence surely with only an approximate form. Nevertheless, the similarity of those scattering cross section at FPs confirms that the concentration and rotation effect of L-type ONM can be effectively achieved by the reduced versions.

## VI. Conclusion.

We revisit ONM from the extension coordinate transformations. Several intriguing devices are designed by L-type ONM. Since the fabrication of L-type ONM is quite challenging. We show that the property of ONM can be effectively achieved by the simplified materials with FPs. Moreover, we propose the reduced versions to construct concentrator and rotator, which only needs dielectric and thin metal plates. The Hz field patterns and total scattering cross sections verify their functionalities. The reduced versions are neither based on metamaterials, nor on photonic crystals. It requires the phase accumulation along the extension paths, hence could bring in a further experimental fabrication of L-type ONM. Such a reduced method can be generalized to manipulate the propagation of other classical waves, and extended to 3D ONM as well.

This work was supported by the National Science Foundation of China for Excellent Young Scientists (Grant No. 61322504), the National Basic Research Program of China (Grant No. 2013CB035901), the Fundamental Research Funds for the Central Universities (Grant No. 20720170015), and the National Science Foundation of China (Grants No. 51779224, No. 51579221, No. 51279180, No.61705200, and No. 11874311).

L. X., Q. W., and Y. Z. contributed equally to this work. L. X. and H. C. conceived the idea and supervised the project. L. X. and Q. W. performed the simulations. All the authors wrote the paper.

*xuin@ahu.edu.cn
†kenyon@xmu.edu.cn

References
[1] U. Leonhardt, Science **312,** 1777 (2006).
[2] J. B. Pendry, D. Schurig, and D. R. Smith, Science **312,** 1780 (2006).
[3] U. Leonhardt, and T. G. Philbin, New J. Phys. **8,** 247 (2006).
[4] U. Leonhardt, and T. Philbin, Geometry and light: the science of invisibility, Dover Inc. Mineola New York, 2010.
[5] A. Einstein, Annalen der Physik **354,** 769 (1916).
[6] H. Chen, C. T. Chan, and P. Sheng, Nat. Mater. **9,** 387 (2010).
[7] A. V. Kildishev, and V. M. Shalaev, Physics-Uspekhi **54,** 53 (2011).
[8] B. Zhang, Light: Science & Applications **1,** e32 (2012).
[9] P. Kinsler, and M. W. McCall, Photonic. Nanostruct. **15,** 10 (2015).
[10] F. Sun, B. Zheng, H. Chen, W. Jiang, S. Guo, Y. Liu, Y. Ma, and S. He, Laser & Photonics Reviews (2017).
[11] M. McCall, J. Pendry, V. Galdi, Y. Lai, S. Horsley, J. Li, J. Zhu, R. Mitchell-Thomas, O. Quevedo-Teruel, and P. Tassin, Jounal of optics **20,** 063001 (2018).
[12] L. Xu, and H. Chen, Nat. Photonics **9,** 15 (2014).
[13] D. Schurig, J. Mock, B. Justice, S. A. Cummer, J. B. Pendry, A. Starr, and D. Smith, Science **314,** 977 (2006).
[14] J. Li, and J. Pendry, Phys. Rev. Lett. **101,** 203901 (2008).
[15] R. Liu, C. Ji, J. J. Mock, J. Y. Chin, T. J. Cui, and D. R. Smith, Science **323,** 366 (2009).
[16] H. F. Ma, and T. J. Cui, Nat. Commun. **1,** 21 (2010).
[17] M. Rahm, D. Schurig, D. A. Roberts, S. A. Cummer, D. R. Smith, and J. B. Pendry, Photonic. Nanostruct. **6,** 87 (2008).
[18] M. M. Sadeghi, S. Li, L. Xu, B. Hou, and H. Chen, Scientific Reports **5,** 8680 (2015).
[19] P. Zhao, L. Xu, G. Cai, N. Liu, and H. Chen, Frontier of physics (2018).
[20] M.-Y. Zhou, L. Xu, L.-C. Zhang, J. Wu, Y.-B. Li, and H.-Y. Chen, Frontiers of Physics **13,** 134101 (2018).
[21] H. Chen, and C. Chan, Appl. Phys. Lett. **90,** 241105 (2007).
[22] H. Chen, and C. Chan, Physical Review B **78,** 054204 (2008).
[23] H. Chen, B. Hou, S. Chen, X. Ao, W. Wen, and C. Chan, Phys. Rev. Lett. **102,** 183903 (2009).

[24] A. Greenleaf, Y. Kurylev, M. Lassas, and G. Uhlmann, Phys. Rev. Lett. **99,** 183901 (2007).

[25] H. Chen, and C. T. Chan, J. Phys. D: Appl. Phys. **43,** 113001 (2010).

[26] C. Li, L. Xu, L. Zhu, S. Zou, Q. H. Liu, Z. Wang, and H. Chen, Phys. Rev. Lett. **121,** 104501 (2018).

[27] H. Chen, J. Yang, J. Zi, and C. T. Chan, EPL **85,** 24004 (2009).

[28] M. Brun, S. Guenneau, and A. B. Movchan, Appl. Phys. Lett. **94,** 061903 (2009).

[29] A. Norris, and A. Shuvalov, Wave Motion **48,** 525 (2011).

[30] S. Narayana, and Y. Sato, Phys. Rev. Lett. **108,** 214303 (2012).

[31] D. Colquitt, M. Brun, M. Gei, A. Movchan, N. Movchan, and I. Jones, Journal of the Mechanics and Physics of Solids **72,** 131 (2014).

[32] H. Xu, X. Shi, F. Gao, H. Sun, and B. Zhang, Phys. Rev. Lett. **112,** 054301 (2014).

[33] T. Han, X. Bai, D. Gao, J. T. Thong, B. Li, and C.-W. Qiu, Phys. Rev. Lett. **112,** 054302 (2014).

[34] X. Huang, Y. Lai, Z. H. Hang, H. Zheng, and C. Chan, Nat. Mater. **10,** 582 (2011).

[35] J. Luo, Y. Yang, Z. Yao, W. Lu, B. Hou, Z. H. Hang, C. Chan, and Y. Lai, Phys. Rev. Lett. **117,** 223901 (2016).

[36] A. Lakhtakia, Int. J. Infrared Millimeter Waves (2002).

[37] I. Liberal, and N. Engheta, Nat. Photonics **11,** 149 (2017).

[38] W. Yan, M. Yan, and M. Qiu, J. Optics **13,** 024005 (2010).

[39] Q. He, S. Xiao, X. Li, and L. Zhou, Opt. Express **21,** 28948 (2013).

[40] J. B. Pendry, Phys. Rev. Lett. **85,** 3966 (2000).

[41] D. R. Smith, W. J. Padilla, D. Vier, S. C. Nemat-Nasser, and S. Schultz, Phys. Rev. Lett. **84,** 4184 (2000).

[42] R. A. Shelby, D. R. Smith, and S. Schultz, Science **292,** 77 (2001).